\documentstyle[epsf]{elsart} 
   
\begin{document} 
\begin{frontmatter} 
 
\title{Chaos and rotational damping in particle-rotor model}  
\author{Javid A. Sheikh$^{1,2}$ and Yang Sun$^{3,4,5}$} 
\address{$^1$
Physik-Department, Technische Universit\"at
M\"unchen, D-85747 Garching, Germany\\
$^2$Department of Physics,
University of Kashmir,
Hazrathbal, Srinagar,
Kashmir, 190 006,
India\\
$^3$Department of Physics, University of Notre Dame, 
Notre Dame, Indiana 46556, U.S.A.\\ 
$^4$ Department of Physics, Tsinghua University, Beijing 100084, P.R. China\\ 
$^5$Department of Physics, Xuzhou Normal University, 
Xuzhou, Jiangsu 221009, P.R. China\\ 
} 
 
\maketitle 
\begin{abstract}

The onset of chaos and the mechanism of rotational damping are studied 
in an exactly soluble particle-rotor model. It is shown that the 
degree of chaoticity as inferred from the statistical 
measures is closely related to the onset of rotational 
damping obtained using the model Hamiltonian. 
 
\end{abstract} 
\end{frontmatter} 
 
 
\section{Introduction}

The study of quantum chaos in many-body problems using the  
random matrix theory (RMT) \cite{meh91} is now an established field of  
research in physics \cite{rei92,gmw98}. It is known that the statistical  
properties of a quantum many-body system which follow the predictions of 
RMT have chaotic behaviour in the classical phase space \cite{bgs84}. This behaviour of 
a quantum many-body system was first observed for the neutron resonance states 
of compound nuclei \cite{hpb82}. It was shown that the fluctuation properties  
of the neutron 
resonance states follow the Gaussian Orthogonal Ensemble (GOE) statistics which is one of 
universal classes of RMT. The statistical analysis of the nuclear states which 
are populated through $\gamma$-decay has also been performed 
\cite{mdhf93,kpr95,skr01,abe90,mdvb93,ym97,zele,gar97,mat97}. It has been 
demonstrated using different theoretical approaches that at low-excitation energies the  
statistical analysis depicts a Poissonian distribution which corresponds to 
underlying regular classical motion. However, at high-excitation energy  
the distribution follows the GOE statistics, implying that  
the underlying classical motion is chaotic. This correlation between the  
statistical distribution and the classical motion has been obtained for a 
large class of physical systems \cite{gmw98}. 
 
However, the important question in the study of quantum chaos, which has remained largely 
unanswered, is: {\it what are the consequences of quantum chaos in  
measurable quantities}? It has recently been demonstrated \cite{skr01} that, in a nuclear system, the  
overlap-integral between the wave functions of the neighbouring rotational 
states can provide a measure of the degree of chaoticity. However, the 
overlap-integral is not directly a measurable quantity. What is needed to be 
calculated is the electromagnetic-transitions  
between the two states. In the previous work \cite{skr01}, the cranking  
mean-field model was used to calculate the overlaps. The cranking wave function  
does not have a well-defined angular momentum, and in principle, is not appropriate to study the  
electromagnetic transition probabilities. 
In order to avoid any errors due to usage of mean-field approximation, in the present work, we have  
developed a particle-rotor model (PRM) approach in which angular momentum  
is strictly conserved. We would like to use the PRM wave functions to study the consequences of the chaotic  
motion on the electromagnetic transitions. 
 
It should be mentioned that the question of rotational damping has also been  
addressed recently by other authors \cite{abe90,mdvb93,ym97} and it has been shown that the phenomenon of  
rotational damping observed in deformed nuclei is a possible  
manifestation of quantum chaos. In these analysis, the rotational bands were obtained 
as the intrinsic excitations of a Cranked-Nilsson mean-field potential. In order 
to obtain the rotational damping, the bands were mixed using a two-body 
residual interaction. Since in the cranked mean-field approaches, as already mentioned, 
angular momentum is not a conserved quantity, these calculations are  
suitable for very high-spin states for which cranking becomes a good  
approximation to the angular momentum projection method.  
 
The purpose of the present work is to present an alternative analysis of the 
rotational damping using the particle-rotor model. Pairing  
is explicitly included in the present calculations and, therefore, the  
results discussed are applicable to all the spin regimes. 
Furthermore, we use the same Hamiltonian for obtaining the bands structures  
as well as the rotational damping. In the earlier studies  
\cite{abe90,mdvb93,ym97},  
the Nilsson potential was used to obtain the intrinsic bands and the rotational  
damping was then calculated with a residual two-body interaction.  
 
The exact analysis in the PRM  
limits the configuration space that can be employed in the numerical studies.  
Therefore, we shall work in a model space of a deformed single-j 
shell as has been done in the earlier studies \cite{kpr95,skr01}. This  
would imply that our 
results cannot be compared directly with the experimental data. The main essence 
of the present work would be to draw a connection between the mechanism 
of rotational damping and the chaotic features in an exactly soluble model. 
 
The manuscript is organized as follows: In the next section, we give some  
relevant details of the PRM. The PRM is a standard model and the 
discussion of this approach can be found in the textbooks, for  
example in Ref. \cite{BMII}. We shall, therefore, only give the  
relevant expressions for the quantities which are used in 
the present work. The results of the calculations are presented  
and discussed in section III. Finally, the present study is  
summarized in section IV.

\section{Particle-rotor model} 
 
The basic philosophy of the PRM \cite{BMII} is to consider the nucleus as a core  
with a few valance particles. In most of the cases, the core is assumed  
to be inert with fixed properties. The Hamiltonian of the PRM is written as the rotor plus the particle part  
 
\begin{equation} 
\hat H = \hat H_{rot} + \hat H_{part}, \label{H1} 
\end{equation} 
with 
\begin{equation} 
\hat H_{rot} = {\frac {\hat R^2} {2 \theta}}. \label{H2} 
\end{equation} 
The quantity $\theta$ in Eq. (\ref{H2}) is the moment of inertia of the core and $\hat R$ is 
the angular momentum of the core. 
In the present work, we shall consider the axially-symmetric core with the  
z-axis 
as the symmetry-axis. $\hat H_{rot}$ in Eq. (\ref{H2}) can be explicitly rewritten as 
\begin{equation} 
\hat H_{rot} = {\frac {(\hat R_1^2+ \hat R_2^2)} {2 \theta}} 
        = {\frac {(\hat I^2 -\hat I_3^2)} {2\theta}}+{\frac {(\hat J^2 - \hat J_3^2)} {2\theta}}  
             - {\frac {(\hat I_+ \hat J_-+\hat I_- \hat J_+)} {2\theta}}, \label{H4}  
\end{equation} 
where $\hat I =\hat R + \hat J$ is the total angular momentum and $\hat J$ is the  
angular momentum of the valance particles. The particle part of the Hamiltonian (\ref{H1}) is given by 
\begin{equation} \label{H} 
\hat H_{part} = -4 \kappa { \sqrt { 4 \pi \over 5 }} \hat Y_{20} -g \delta(\hat r_1 - \hat r_2). 
\label{residual}
\end{equation} 

The strength of the two-body interaction is 
$G=g\int R^4_{nl}r^2 dr$. It
should be noted that the delta-interaction used in the
present work has predominant pairing properties. However,
since we solve the two-body interaction exactly, 
it will contribute to both pairing or particle-particle
channel as well as to the particle-hole channel. The deformation 
energy $\kappa$ is related to the deformation parameter $\beta$ through \cite{swi98} 
\begin{equation} 
\kappa \simeq 0.16\hbar\omega_0(N+3/2)\beta, 
\end{equation} 
where $\hbar\omega_0$ is the harmonic oscillator frequency of the 
deformed potential and $N$ the quantum number of the major shell.  
For the case of $i_{13/2}$ shell, $\kappa$=2.5 approximately 
corresponds to $\beta=0.28$.  
 
The wave function of the PRM is given by  
\begin{equation} 
\left| \Psi _{IM} \right> = \sum_{K_i} c_{K_i} \left| IMK_i \right> \label{w1}, 
\end{equation} 
where 
\begin{equation} 
\left|IMK_i\right> = \sqrt{ \frac {2I+1} {16\pi^2 (1+\delta_{K0})}} 
\biggr\{ D^{*I}_{MK} \left|\phi_{K_i}\right> + (-1) ^I 
e^{-\imath \pi \hat J_1} D^{*I}_{M-K} \left|\phi_{K_i}\right> \biggr\} .
\end{equation} 
The wave function $\left|\phi_{K_i}\right>$ is the solution of the intrinsic particle Hamiltonian, $H_{part}$. 
For $K=0$, these states have well-defined symmetry with respect to rotation about the x-axis, 
$e^{-\imath \pi \hat J_1}$.  
 
The wave function obtained by diagonalizing the Hamiltonian, Eq. (\ref{H1}), is used  
to evaluate the electromagnetic transition probabilities.  
The reduced electric transition probability B(EL) from an initial state  
$( I_i ) $ to a final state $(I_f )$ is given by  
\begin{equation} 
B(EL,I_i \rightarrow I_f) = {\frac {1} {2 I_i + 1}}  
|\left< \Psi_{I_f} || \hat Q_L || \Psi_{I_i} \right>|^2 , 
\label{be2} 
\end{equation} 
where the reduced matrix element can be expressed as   
\begin{eqnarray} 
&&\left< \Psi_{I_f} || \hat Q_L || \Psi_{I_i} \right> 
\nonumber \\  
&=& \sqrt{(2 I_i+1)} \sum_{K_i,K_f} c_{I_i K_i} \, c_{I_f K_f} \biggr\{ 
{\frac {1} { \sqrt{(1+\delta_{K_i0})(1+\delta_{K_f0})}}}\nonumber\\ 
&\times&\sum_{q'} \biggr[ 
\left[ 
\begin{array}{ccc} 
I_i & L & I_f \\ 
K_i & q'& K_f 
\end{array} \right] \left<\phi_{K_f}|\hat Q^L_{q'}{\rm{(int)}}|\phi_{K_i}\right> 
\nonumber \\ 
&+& (-1)^{I_i} 
\left[ 
\begin{array}{ccc} 
I_i & L & I_f \\ 
-K_i & q'& K_f  
\end{array} \right] \left<\phi_{K_f}|\hat Q^L_{q'}{\rm{(int)}}|e^{-\imath \pi \hat J_1}\phi_{K_i}\right>  
\biggr]\nonumber\\ 
&+& \delta_{K_i K_f} Q_0  
\left[ 
\begin{array}{ccc} 
I_i & L & I_f \\ 
K_i & 0 & K_f  
\end{array} \right] \biggr\}.  
\label{rme} 
\end{eqnarray} 
The quantity $\left<\phi_{K_f}|\hat Q^L_{q'}{\rm{(int)}}|\phi_{K_i}\right>$ is the matrix-element 
of the quadrupole operator in the intrinsic frame, which can be readily calculated using 
the wave function of the particle Hamiltonian $\hat H_{part}$. $Q_0$ in Eq. (\ref{rme}) 
is the quadrupole moment of the core. 
 
\section{Results and discussions} 
 
The PRM calculations have been performed for six  
valance particles with the standard parameters : 
$G=0.45$ MeV and $\theta=24~ \hbar^2$ MeV$^{-1}$. These parameters have been used in most 
of the earlier studies and are considered to be reasonable for the $j=13/2$  
model space \cite{ak80}. In order to study the deformation dependence, the PRM   
calculations have been done with two sets of deformation values, $\kappa=2.5$ 
and 6.0 MeV. These deformation values correspond to the normal and superdeformed 
shapes. The results of the statistical analysis and the rotational damping are 
presented in the following subsections. 
 
\subsection{Statistical analysis} 
 
The spectral rigidity statistics of Dyson and Mehta \cite{meh91},  
$\bar\Delta_3(L)$, measures the long-range  
correlation of the unfolded levels 
\begin{equation} 
\Delta_3(X,L)=\min {{1}\over{L}}\int ^{X+L}_X [N_u(E)-(AE+B)]^2dE, 
\end{equation} 
where $N_u$ is the cumulative level density of the unfolded 
levels $X_i$. We average $\Delta_3(X,L)$ over intervals $(X,X+L)$ to 
obtain $\bar\Delta_3(L)$, in a way as outlined in Refs. \cite {bgs84,bro81}. 
For ordered systems 
$\bar\Delta_3(L)=L/15$, and for fully chaotic ones  
$\bar\Delta_3(L)\approx 
\ln(L)/\pi^2-3/4$ for $L\gg 1$. 
The explicit calculation of the spectral rigidity was done with 
the method described in Ref. \cite{bo84b}. 
 
 
In Figs. 1 and 2, the results of spectral statistics 
are respectively presented for the two deformation values of $\kappa = 2.5$ and 6.0 MeV. 
The results with all the eigen-values are shown by filled  
circles. In order to study the degree of chaoticity as  
a function of excitation-energy, the statistical analysis has also been  
studied for three more sets of eigen-values. Instead of choosing the energy window, 
we have considered the number  
of eigen-values, since 
the number of states in a given energy window change as a function of angular momentum 
and deformation. Furthermore, in order to have a reasonable statistics we have 
chosen 400 eigen-values in each bin. The results of the  
analysis with eigen-values ranging from 1 to 400 are shown by upper triangles 
and the third set with eigen-values from 401 to 800 are shown by  
lower triangles in Figs. 1 and 2.  We also performed the analysis of the 
eigen-values ranging from 801 to 1200, but the results turn out to be very similar 
to the set of 401 to 800 and are not shown in the figures. 

For the results for $\kappa=2.5$ MeV, corresponding to normal deformation, 
Fig. 1 indicates that at low-spin values 
of I = 10 and 12, $\Delta_3$ values closely follow the curve expected  
for the GOE distribution. It is noted that the GOE statistics is obtained at  
all excitation energies. Therefore, the classical motion is fully chaotic for 
low-spin values. For high angular momenta close to I = 60, the results  
with all the eigen-values deviate from the GOE curve and, 
therefore, indicating that the underlying classical motion is less chaotic.  
However, the results also show that the motion becomes more chaotic at higher 
excitation energies. The $\Delta_3$ values for the set 401-801 eigen-values, 
shown by lower triangles, are more close to the curve expected for the chaotic 
motion. The generic property that the system tends to become regular with  
increasing angular momenta is a consequence \cite{kpr95} of the decoupling      
of the particle spins from the deformation axis of the rotor. 
As the nuclear system rotates faster, more and more particles tend to align their spins 
along the same direction of the rotation axis, which causes the system to  
become more regular. The results for the intermediate spin values of I = 30 and 
32 appear slightly less chaotic as compared to I = 10 and 12, when considering  
all the eigen-values, but are similar to low-spin values at higher excitation 
energies. 
 
In Fig. 2, the results for $\kappa=6.0$ MeV, corresponding to superdeformation, indicate that for the low spin 
values of I = 10 and 12, the motion is less chaotic as compared to Fig. 1.  
At low-excitation energy, the results appear significantly less chaotic, but 
at higher excitation energy the motion tends to become chaotic. 
At higher spins, the results appear to be similar to those of Fig. 1. 
 
\subsection{Distribution of E2 transitions} 
 
At low excitation energies, the E2 transition from a given state populates one 
particular state at a lower-spin. However, at high-excitation energies, 
it is known that a given state populates several states. 
This fragmentation of the E2 strength is referred to as the rotational  
damping. It has 
been shown using the cranking model that the onset of rotational damping at a given 
excitation energy is related to the underlying chaotic  
motion \cite{abe90,mdvb93,ym97}. In the present work, 
we shall also try to investigate this relationship.  
 
In order to evaluate rotational damping, the B(E2) values have been calculated 
for a range of initial and final states. We considered three sets of initial 
states, which represent the three energy sets considered in the statistical 
analysis of the eigen-values in Figs. 1 and 2. The three sets of initial states 
considered are : (a) from 1 to 50, (b) from 275 to 325 and (c) from 575 to 625. 
The decay of each of these initial states has been calculated to all the final 
states at a lower-spin. In Figs. 3 and 4, the results of B(E2) values, for the 
normal deformation value of $\kappa=2.5$ MeV, are plotted 
for three initial states of 26, 300 and 600 as a representative examples 
of the above three chosen sets. In the evaluation of the B(E2) value from 
Eq. (\ref{rme}), the quadrupole-moment of the core, $Q_0$, needs to be specified.  
In order to study the core dependence of the rotational damping, two sets 
of calculations have been carried out - one with no core quadrupole-moment 
and the other with a $Q_0 = 2.5$ barn. The results of these two sets are 
shown in Figs. 3 and 4, respectively.  
 
The results of the B(E2) for $i=26$ depict a peak value at around the final 
state $f=26$, with very small contribution from the neighboring final states. 
For the initial state of $i=300$, the B(E2) is very fragmented; there are many 
states to which this state decays into. For the initial state of $i=600$, the 
fragmentation is even higher. On comparing the results of three transitions 
of $I_i=12 \rightarrow I_f=10$, $I_i=32 \rightarrow I_f=30$ and  
$I_i=62 \rightarrow I_f=60$ in Fig. 1, it is noted that the degree  
of fragmentation decreases with increasing angular momentum.  
 
The B(E2) results with a core quadrupole moment of $Q_0=2.5$ barn, shown 
in Fig. 4, depict a substantially reduced fragmentation as compared  
to Fig. 3. The B(E2) value peaks for one particular transition for which 
the initial state is same as that of the final state. This can be easily 
understood from Eq. (\ref{rme}), since for the core contribution only the diagonal 
term for which $K_i=K_f$ contributes. The fragmentation arises from the terms 
of the intrinsic quadrupole operator ($Q_{\pm 1}, Q_{\pm 2}$) of Eq. (\ref{rme}), which  
mix different K-components in the intrinsic wave function. 
 
The results for $\kappa=6.0$ MeV are shown in 
Figs. 5 and 6. Fig. 5 presents the results with no core quadrupole 
moment and Fig. 6 depicts the results with $Q_0=5$ barn. At low excitation 
energies, it is noted from Fig. 5 that the fragmentation is reduced as compared to Fig. 3, 
but at higher excitation energies the results for the two deformation 
values are very similar. This observation is consistent with that obtained 
with the statistical analysis. The results of Fig. 6 with core quadrupole 
moment, $Q_0=5$ barn, show that fragmentation is substantially reduced for most of the cases. 

Chaos is expected to set in when 
the average level distance is of the order of the coupling matrix elements \cite{abe90}.
In a model study with a limited model space and effective interactions which depend on parameters, 
it is instructive to study the 
density of states as a function of excitation energy and angular momentum, and the interplay 
of the density of states and the residual interactions in the determination of the chaotic motion. 

In Fig. 7, the density of states is plotted by counting energy levels in an energy window of 1 MeV for each
excitation energy. All the levels generated in the present model space are included, without any truncation. 
The density is shown for three angular momenta $I =$ 10, 30, and 60, 
and for the two deformation values of $\kappa = 2.5$ and 6.0 MeV. 
The curves for the normally deformed case with $\kappa = 2.5$ 
show a significantly decreasing trend when angular momentum increases. 
For the superdeformed case with $\kappa = 6.0$, the decreasing trend is not as pronounced as 
the normal deformed case. However, a drop at $I=60$ is still clearly seen. 
Comparing the results for the two deformations, we see that for the superdeformed case, 
the density of states is much smaller. This is particularly true for $I=$ 10 and 30.

If the density of states 
decreases with deformation, a larger residual interaction would be needed to 
obtain a similar amount of fragmentation in E2 strength.
It is thus interesting to investigate situations in the superdeformed case, but with enhanced 
interactions.
In Figs. 8 and 9, the same calculations presented in Figs. 5 and 6 are repeated by using 
a larger residual interaction. The interaction strength $g$ in Eq. (\ref{residual}) is enhanced 
by a factor of 2. It can be seen that with the enhanced interaction,
a stronger fragmentation is now obtained.  
The superdeformed results with enhanced 
interaction (Figs. 8 and 9) seem to approach the normally deformed results with the
ordinary interaction (Figs. 3 and 4). 

\subsection{Rotational damping} 
 
Rotational damping  
is measured through the branching parameter $n$ \cite{mdvb93} 
\begin{equation} 
n = \{ \sum_j [B(E2,I_i \rightarrow I_j)]^2 \}^{-1}, 
\label{bpn}
\end{equation} 
where $B(E2,I_i \rightarrow I_j)$ transition probability is defined in 
Eqs. (\ref{be2}) and (9). In order to evaluate $n$, it is required to calculate  
the B(E2) transition from one particular state to all the states which are 
allowed by the selection rules of the $\gamma-$ray radiation. 
Considering a case where it decays to only one state  
(as is true at low-excitation energy),  
$n$ is equal to one if the B(E2)'s are normalized. For the  
case where a given state decays 
to two possible states with equal probabilities, $n$ is equal to 2. This 
implies that for $n$ less than 2, the band structure  
is discrete and each given state 
decays only to one state at a lower excitation energy. $n$  
greater than 2 corresponds to rotational damping, signifying that B(E2) transition  
strength from a given state populates two or more states at a lower spin. 
 
The branching parameter $n$ for $\kappa=2.5$ MeV is presented  
in Table I for a range of initial states. In order to avoid local fluctuations, 
the $n$ value has been calculated by averaging over the neighboring states. 
In Table I, the number $1-10$ in the first column, for example, indicates that $n$ 
has been calculated by averaging over the initial eigen-states from 1 to 10. 
Of course, all possible final states have been used to calculate $n$ 
from each initial state before averaging. 
 
The second column of Table I gives the results of $n$ for the 
transition $I_i=12 \rightarrow I_f=10$. It is observed from Table I that  
$n$ with no core quadrupole moment becomes greater than 2 even at 
a very low excitation energy and increases with increasing excitation energy. 
For the case with a core quadrupole moment, $Q_0=2.5$ barn, there is no 
damping at low-excitation energy and  $n$ increases very slowly 
as compared to the case with no quadrupole moment. This suppression of the 
rotational damping with a core contribution indicates that the damping is 
essentially determined by the contribution of valence particles. 
 
For the second set of initial states from $i = 275$ to 324, the value of $n$ 
has now increased to more than 100 and for the third set of states from 
$i = 575$ to 625, $n$ has increased further. In general, we note that 
$n$ increases with increasing excitation energy. However, for the 
case $615-625$, $n$ has in fact decreased. This drop in $n$ 
can be understood from the fact that the present model space is finite, and 
as we reach the top of the shell, the level density decreases \cite{kpr95}. 
In a realistic model, the level density should increase with increasing excitation energy.  
 
In the third column of Table I, $n$ is given for the transition 
$ I_i=32 \rightarrow I_f=30 $. $n$ is now reduced as compared to the 
transition  $ I_i=12 \rightarrow I_f=10 $. In the last column, the results 
are presented for the transition  $ I_i=62 \rightarrow I_f=60$. The numbers 
are now considerably reduced. For the results with a core quadrupole moment, 
$Q_0=2.5$, rotational damping is completely absent at low excitation energy, 
$n$ is less than 2 for almost all the first set of states. 
 
The results of the rotational damping for $\kappa$ = 6 MeV are  
presented in Table II for the 
three transitions as studied for $\kappa$ = 2.5 MeV. For the  
transition $I_i=12 \rightarrow I_f=10$ 
the damping is observed at a higher excitation energy. For the transitions  
$ I_i=32 \rightarrow I_f=30 $ and $ I_i=62 \rightarrow I_f=60 $, the results appear to be similar to Table I. 
These findings are consistent with the conclusion drawn in the previous figures. 
Our calculations with enhanced interaction strength for $\kappa$ = 6 MeV clearly indicate 
an increased damping. For example, the branching numbers $n$ averaged for the initial eigen-states
from 41 to 50 in the transition of $I_i = 62 \rightarrow I_f = 60$ were 6.83 when $Q_0=0$, and 1.50 when 
$Q_0=5.0$ (see Table II). With enhanced interaction, they increase to 8.20 and 1.65, respectively.

\section{Summary and conclusions} 
 
In the present study, we tried to address a fundamental issue on the {\it observable  
consequences of quantum chaos in rotating nuclei}. It is now well accepted that the statistical analysis 
of a quantum many-body system can provide the important information on the underlying 
motion. In case that the statistical distribution follows GOE statistics, the underlying 
motion is considered to be chaotic, and for the case where the statistical distribution  
is Poissonian, the underlying motion is regarded to be regular. This statistical 
distribution has been studied for a wide range of physical systems. For the 
case of rotating nuclei, it is known that at low-excitation energy, the motion 
is regular, but with increasing excitation energy the motion tends to become 
chaotic. The angular momentum dependence of the statistical distribution has 
also be studied and it has been shown that for low angular momenta, the  
distribution follows GOE and for higher angular momenta the distribution is 
Poissonian.

The important issue that needs to be explored is what observable 
quantities will be sensitive to the changes in the underlying motion of a physical 
system. It has been recently shown that the phenomenon of rotational 
damping observed in rotating nuclei may provide a valuable information on the 
underlying motion. In the present work, we have investigated the rotational 
damping and the statistical distribution for measurable quantities in an exactly soluble model  
with conserved angular momentum. We consider that 
it is important to make an exact study since any approximation may lead to different 
errors in the calculation of the rotational damping and the statistical distributions.  
Consequently, it may not be possible to make a comparison of the two 
calculated quantities in an accurate manner. 
 
To summarize, in the present work, we have employed the particle-rotor model with particles in 
a deformed single-j shell of $i_{13/2}$, which interact through a two-body  
delta-interaction. The model has been solved exactly with different input 
parameters. Within this model space,  
the present analysis has shown a clear correspondence between 
the mechanism of rotational damping and the onset of chaos in rotating nuclei.  
In particular, we have observed the following similarities: 
\begin{enumerate} 
\item 
increase in rotational damping with excitation-energy and the corresponding increase in 
the degree of chaoticity; 
\item 
decrease in rotational damping with increasing angular momentum and the corresponding decrease 
in the degree of chaoticity;  
\item 
decrease in rotational damping with increasing deformation and the corresponding decrease in the  
degree of chaoticity.  
\end{enumerate} 
 
Research at the University of Notre Dame is supported by the NSF under contract number PHY-0140324.

\newpage 
 
\begin{table}[h] 
\begin{center} 
\caption{ The results of the branching number $n$, defined in Eq. (\ref{bpn}), for the 
normal deformation case with $\kappa = 2.5$ MeV. The numbers presented are averaged for 
ten initial states. For example, the number $1-10$ in the first column signifies 
that the values have been averaged for initial eigen-states from 1 to 10. 
             } 
\begin{tabular}{|c|rl|rl|rl|} 
\hline  
Initial  & $ I_i = 12$ & $ \rightarrow I_f = 10$ 
& $ I_i = 32$ & $ \rightarrow I_f = 30$  
& $ I_i = 62$ & $ \rightarrow I_f = 60$ \\ 
eigenstates  & $n_{(Q_0 =0)}$ & $n_{(Q_0=2.5)}$  
             & $n_{(Q_0 =0)}$ & $n_{(Q_0=2.5)}$  
             & $n_{(Q_0 =0)}$ & $n_{(Q_0=2.5)}$ \\ \hline 
$  1- 10$&$   2.09$&$   1.48$&$   2.56$&$   1.35$&$   4.43$  &$   1.54$  \\ 
$ 11- 20$&$   6.96$&$   3.75$&$   5.81$&$   2.41$&$   7.33$  &$   1.92$  \\  
$ 21- 30$&$   9.79$&$   5.43$&$   8.16$&$   2.76$&$   7.02$  &$   2.02$  \\  
$ 31- 40$&$  11.77$&$   5.51$&$   9.68$&$   2.54$&$   9.62$  &$   1.83$  \\  
$ 41- 50$&$  17.41$&$   8.47$&$  14.70$&$   3.78$&$   9.38$  &$   1.91$  \\ 

\hline
$275-284$&$ 107.16$&$  35.52$&$  70.18$&$   7.18$&$  21.91$  &$   4.54$  \\ 
$285-294$&$ 102.19$&$  34.24$&$  61.52$&$   8.92$&$  24.38$  &$   3.77$  \\ 
$295-304$&$ 108.24$&$  34.05$&$  69.33$&$  11.56$&$  22.36$  &$   3.29$  \\ 
$305-314$&$ 110.05$&$  35.48$&$  64.53$&$   7.94$&$  18.37$  &$   4.47$  \\ 
$315-324$&$ 115.78$&$  39.24$&$  65.83$&$   7.78$&$  23.89$  &$   4.48$  \\  
\hline
$575-584$&$ 140.82$&$  50.96$&$  79.99$&$   9.79$&$  21.67$  &$   5.60$  \\ 
$585-594$&$ 145.62$&$  50.71$&$  73.37$&$   8.18$&$  19.22$  &$   2.89$  \\ 
$595-604$&$ 137.70$&$  48.95$&$  63.17$&$   8.36$&$  13.18$  &$   3.20$  \\ 
$605-614$&$ 149.51$&$  54.33$&$  71.78$&$  11.52$&$  25.54$  &$   2.75$  \\ 
$615-624$&$ 146.49$&$  56.60$&$  64.41$&$   9.37$&$  22.53$  &$   3.93$  \\  
\hline  
\end{tabular} 
\end{center} 
\end{table} 
 
\newpage 
\begin{table}[h] 
\begin{center} 
\caption{ The same as Table I, but for the superdeformation case with 
$\kappa = 6$ MeV. 
            } 
\begin{tabular}{|c|rl|rl|rl|} 
\hline  
Initial  & $ I_i = 12$ & $ \rightarrow I_f = 10$ 
& $ I_i = 32$ & $ \rightarrow I_f = 30$  
& $ I_i = 62$ & $ \rightarrow I_f = 60$ \\ 
eigenstates  & $n_{(Q_0 =0)}$ & $n_{(Q_0=5.0)}$  
             & $n_{(Q_0 =0)}$ & $n_{(Q_0=5.0)}$  
             & $n_{(Q_0 =0)}$ & $n_{(Q_0=5.0)}$ \\ \hline 
$  1- 10$  &$   1.91$  &$   1.34$&$   1.68$  &$   1.12$&$   2.57$  &$   1.06$ \\ 
$ 11- 20$  &$   2.40$  &$   1.55$&$   2.74$  &$   1.49$&$   4.61$  &$   1.36$ \\ 
$ 21- 30$  &$   5.33$  &$   2.67$&$   3.51$  &$   1.69$&$   6.18$  &$   1.65$ \\ 
$ 31- 40$  &$   7.37$  &$   3.50$&$   5.13$  &$   2.36$&$   6.77$  &$   1.39$ \\ 
$ 41- 50$  &$   8.23$  &$   3.58$&$   5.82$  &$   2.17$&$   6.83$  &$   1.50$ \\ 
\hline
$275-284$  &$  57.52$  &$  15.52$&$  51.89$  &$   8.42$&$  25.89$  &$   2.53$ \\ 
$285-294$  &$  47.96$  &$  10.07$&$  44.85$  &$   7.67$&$  29.28$  &$   2.35$ \\ 
$295-304$  &$  54.73$  &$  11.61$&$  63.29$  &$   9.56$&$  27.50$  &$   3.03$ \\ 
$305-314$  &$  59.36$  &$  10.36$&$  63.47$  &$  11.21$&$  40.92$  &$   4.02$ \\ 
$315-324$  &$  60.90$  &$  14.61$&$  58.97$  &$  10.12$&$  36.44$  &$   3.19$ \\ 
\hline
$575-584$  &$  98.83$  &$  16.41$&$  82.31$  &$  13.48$&$  40.13$  &$   3.25$ \\ 
$585-594$  &$ 106.70$  &$  20.00$&$  96.13$  &$  12.38$&$  29.94$  &$   2.45$ \\ 
$595-604$  &$ 119.50$  &$  14.66$&$  91.19$  &$  10.37$&$  35.23$  &$   2.97$ \\ 
$605-614$  &$  95.99$  &$  14.59$&$  65.22$  &$  10.09$&$  34.68$  &$   2.69$ \\ 
$615-624$  &$ 100.95$  &$  13.55$&$  75.04$  &$  11.68$&$  23.36$  &$   2.14$ \\  
\hline  
\end{tabular} 
\end{center} 
\end{table} 
 
\newpage 
\begin{figure} 
\caption{ 
The spectral-rigidity parameter of Dyson and Mehta for $\kappa=2.5$ MeV 
for various angular momenta. Three sets of calculations have been carried out. 
In the first set (shown by filled circle), the  
analysis have been done with all the eigen-values, in the second case 
(shown by upper triangle) the eigen-values ranging from 1 to 400 have been 
used, and in the third case (shown by lower triangle)  
the eigen-values from 401 to 800 have been selected. 
} 
\label{figure.1} 
\end{figure} 
 
\begin{figure} 
\caption{ 
The same as in Figure 1, but for $\kappa=6$ MeV.
} 
\label{figure.2} 
\end{figure} 
 
\begin{figure} 
\caption{ 
The normalised B(E2) transition probabilities obtained 
for $\kappa$ = 2.5 MeV with no core quadrupole moment, $Q_0=0$.  
The results are shown for the transitions (a)  
$ I_i=12 \rightarrow I_f=10 $, (b) $ I_i=32 \rightarrow I_f=30 $ and  
(c) $ I_i=62 \rightarrow I_f=60 $. For each case, the transition probability 
is given for three different initial eigen-states of $i$ = 26, 300 and 600. 
} 
\label{figure.3} 
\end{figure} 
 
\begin{figure} 
\caption{ 
The same as in Figure 3, but 
for $\kappa$ = 2.5 MeV with core quadrupole moment, $Q_0=2.5$ barn.  
} 
\label{figure.4} 
\end{figure} 
 
\begin{figure} 
\caption{ 
The same as in Figure 3, but 
for $\kappa$ = 6.0 MeV with no core quadrupole moment, $Q_0=0$.  
} 
\label{figure.5} 
\end{figure} 
 
\begin{figure} 
\caption{ 
The same as in Figure 3, but 
for $\kappa$ = 6.0 MeV with core quadrupole moment, $Q_0=5.0$ barn.  
} 
\label{figure.6} 
\end{figure} 

\begin{figure}
\caption{
Density of states as a function of excitation energy. Calculations are
performed for three representative angular momenta $I= $ 10, 30, and 60,
and for the two deformations $\kappa$ = 2.5 and 6 MeV.  
} 
\label{figure.7}
\end{figure}

\begin{figure}
\caption{
The same as in Figure 5, but with enhanced residual interaction. 
} 
\label{figure.8}
\end{figure}

\begin{figure}
\caption{
The same as in Figure 6, but with enhanced residual interaction. 
} 
\label{figure.9}
\end{figure}

\end{document}